\documentclass[lettersize,journal]{IEEEtran}
\usepackage{amsmath,amsfonts}
\usepackage{algorithmic}
\usepackage{algorithm}
\usepackage{array}
\usepackage[caption=false,font=normalsize,labelfont=sf,textfont=sf]{subfig}
\usepackage{textcomp}
\usepackage{stfloats}
\usepackage{url}
\usepackage{verbatim}
\usepackage{graphicx}
\usepackage{cite}
\usepackage{xcolor}
\usepackage{tabularx}
\usepackage{booktabs}
\usepackage{bbm}
\usepackage[flushleft]{threeparttable}

\begin{document}
\bstctlcite{IEEEexample:BSTcontrol}

\title{On the Role of ViT and CNN in Semantic Communications: Analysis and Prototype Validation}

\author{Hanju Yoo,~\IEEEmembership{Graduate Student Member,~IEEE,}
Linglong Dai,~\IEEEmembership{Fellow,~IEEE,}\\
Songkuk Kim,~\IEEEmembership{Member,~IEEE,}
and Chan-Byoung Chae,~\IEEEmembership{Fellow,~IEEE}
\thanks{H.~Yoo, S.~Kim, and C.-B.~Chae are with Yonsei University, Korea. Email: \{hanju.yoo, songkuk, cbchae\}@yonsei.ac.kr. L.~Dai is with Tsinghua University, Beijing 100084, China. Email: daill@tsinghua.edu.}
}

\markboth{}%
{Yoo \MakeLowercase{\textit{et al.}}: On the Role of ViT and CNN in Semantic Communications: Analysis and Prototype Validation}

\IEEEpubid{}

\maketitle

\begin{abstract}
Semantic communications have shown promising advancements by optimizing source and channel coding jointly. However, the dynamics of these systems remain understudied, limiting research and performance gains. Inspired by the robustness of Vision Transformers (ViTs) in handling image nuisances, we propose a ViT-based model for semantic communications. Our approach achieves a peak signal-to-noise ratio (PSNR) gain of +0.5 dB over convolutional neural network variants. We introduce novel measures, average cosine similarity and Fourier analysis, to analyze the inner workings of semantic communications and optimize the system's performance. We also validate our approach through a real wireless channel prototype using software-defined radio (SDR). To the best of our knowledge, this is the first investigation of the fundamental workings of a semantic communications system, accompanied by the pioneering hardware implementation. To facilitate reproducibility and encourage further research, we provide open-source code, including neural network implementations and LabVIEW codes for SDR-based wireless transmission systems.\footnote{Source codes available at https://bit.ly/SemViT}
\end{abstract}


\section{Introduction}


Conventional communications systems traditionally employ separate blocks for source coding and channel coding. This modular approach stems from Shannon's separation theorem~\cite{shannon1948mathematical}, which asserts that source coding and channel coding can be independently optimized without compromising optimality, under idealized communication conditions like infinite code length, independent and identically distributed (i.i.d.) symbols, or stationary channels. However, in practical communication scenarios, these assumptions often do not hold~\cite{vembu1995source}.

To bridge the gap between theoretical assumptions and real-world conditions, there is growing interest in semantic communications. Semantic communications aims to address these challenges by integrating source coding, channel coding, and modulation within a joint optimization framework based on deep learning techniques~\cite{BeyondBits_all, bourtsoulatze2019deep, farsad2018deep}. By considering the interplay between these components, semantic communications holds promise for achieving improved transmission performance and enhanced efficiency in real-world communication systems.

Typically, these systems treat wireless channels as a non-trainable, noise-adding layer, and are trained end-to-end~\cite{bourtsoulatze2019deep, xie2020lite,  farsad2018deep, kurka2020deepjscc, dai2022nonlinear}. Because neural networks consist of various matrix operations, they are highly parallelizable and can be scaled to meet performance and latency requirements, which is critical in emerging applications such as extended reality (XR)~\cite{liaskos2022xr}. Unlike traditional coding methods that aim to recover the original symbols precisely, deep-learning-based models can be optimized for goal-oriented communications. For example, learned representations can target more accurate classifications of an image~\cite{strinati20216g} rather than full signal recovery, which may not be necessary.

\IEEEpubidadjcol

The semantic communications system can be viewed as a deep neural network problem that involves reconstructing the original signal from the corrupted latent representations. One of the main research focuses is selecting building blocks for the neural network, such as convolutional~\cite{bourtsoulatze2019deep, kurka2020deepjscc}, self-attention~\cite{xie2020lite, dai2022nonlinear}, or recurrent neural network blocks~\cite{farsad2018deep}, or finding optimal network architecture. While heuristics can be used to find optimal architectures, a deeper analysis of how deep neural networks jointly perform source/channel coding or modulations, which is currently understudied, can significantly enhance these architecture search procedures and facilitate research.

In this paper, we extend our prior work~\cite{yoo2022real}, which merely adopted the Vision Transformer (ViT) architecture to semantic communications systems, and carefully fine-tune the network to find better architectures for the system. Moreover, we thoroughly analyze the results to understand how the image semantic communications systems work and what the advantage of the ViT is. We also verify the results in real wireless channels using a software-defined-radio (SDR)-based testbed. Full source codes is available at https://bit.ly/SemViT, including SDR implementations and trained neural network parameters. Our contributions are as follows:

\begin{itemize}
\item We carefully design semantic communications systems that harmonize Vision Transformers with CNNs, with appropriate priors and insights from the computer vision community.
\item We conduct extensive analysis about how image semantic communications systems work in an additive white Gaussian noise (AWGN) or Rayleigh channel, yielding various insights while introducing analysis metrics such as average cosine similarities and Fourier analysis.
\item We build a SDR-based wireless semantic communications system prototype and verify that the simulated results fit well with the real wireless channels.
\item We publicly release our source codes, including deep neural network implementations, trained parameters, or SDR-based testbed implementations, to facilitate follow-up studies.
\end{itemize}

\section{Backgrounds} \label{backgrounds}

\subsection{Semantic Communications} \label{intro_semantic}

A semantic communications system is a kind of autoencoder, a neural network capable of compressing given signals and reconstructing them. A typical autoencoder encodes the given image into the smaller but high-dimensional features and then reversely generates the output of the latent representations to match the original signals. In that process of squeezing and expanding, only the core part of the input is extracted in an unsupervised manner~\cite{BeyondBits, BeyondBits_all}. One way to exploit those properties is a denoising autoencoder~\cite{vincent2008extracting}, which removes artifacts in a noised image.

The semantic communication system, however, may be distinguished from the denoising autoencoder regarding the location of the noise. In the denoising autoencoder, the original image contains the artifacts. In the contrary, in the semantic communications system, perturbations are added to the compressed latent features (i.e., encoded symbols). Those perturbations typically consist of theoretical Rayleigh fading or AWGN. The networks learn to generate compressed, noise-resilient signal representations by using loss functions that measure the distance between the original signals and the reconstructed ones (e.g., mean-squared error loss). If those learned features are composed of two-dimensional complex numbers, those autoencoder networks can be viewed as a mapping function from the source to a symbol. Unlike traditional communications systems that separate source coding, channel coding, and modulation, a semantic communications system conducts joint source-channel coding using a deep neural network.

Ongoing research on semantic communications covers various domains, including texts~\cite{farsad2018deep, xie2020lite, xie2021deep}, speech signals~\cite{weng2021semantic}, images~\cite{bourtsoulatze2019deep, kurka2020deepjscc}, and videos~\cite{tung2021deepwive}. For example, the authors in \cite{xie2021deep} proposed a transformer-based semantic communications architecture, DeepSC. \cite{weng2021semantic} expanded the domain to speech transmission with squeeze-and-excitation-aided convolutional neural networks (CNN). The work \cite{bourtsoulatze2019deep} proposed a joint source and channel coding (JSCC) network for images, which compresses given images with CNNs. Their coding method achieved 3~dB peak signal-to-noise-ratio (PSNR) gain to JPEG+LDPC coding scheme under the Rayleigh fading channel. \cite{tung2021deepwive} presented DeepWiVe, an video transmission JSCC system that uses CNN and non-local blocks~\cite{wang2018non} to capture the redundancies between frames.

However, most prior work mentioned above has the following limitations: 1) they mostly have focused on the implementation rather than delicate considerations of the network architecture or careful analysis of why the proposed system works; 2) they have evaluated the system only in the simulated channel environments (i.e., Rayleigh and AWGN channels). In this paper, we propose a ViT-based image semantic communication system inspired by existing analysis of the computer vision communities on ViTs; we carefully analyze the results with various metrics to figure out how the ViTs work in the semantics communications system. We also verify the system with the SDR-based wireless prototype to show the system's feasibility in a real wireless channel.

Recently, there have been a few pioneering works that have used Vision Transformer-based backbones for semantic communications. For example, the authors in \cite{hu2022robust} used a ViT encoder to produce semantic-noise resilient features inspired by ViT's patch-wise image processing. The authors in \cite{dai2022nonlinear} used a Swin Transformer-based backbone for image analysis/transform to bring the hyperprior concept to the semantic communications.

However, they used ViTs only to realize their ideas, while we aim here--through extensive analysis--to figure out how the ViT works in sematic communications via extensive analysis and try to provide some insights. We also verify our results on a real wireless channel with our SDR-based testbed. Our contributions are parallel to those works and can be utilized simultaneously to improve communications systems.

\begin{figure*}[htbp]
\centering{\includegraphics[width=\textwidth]{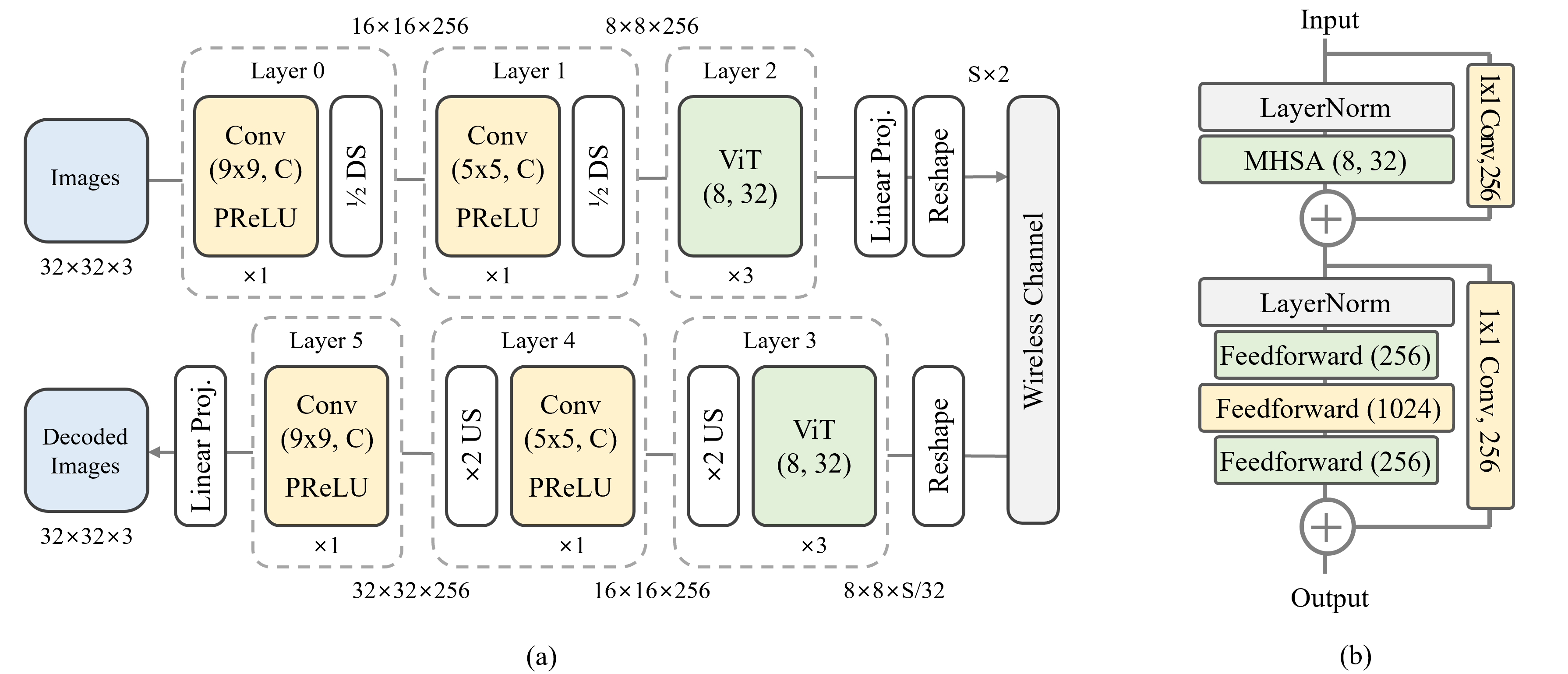}}
\caption{(a) Proposed system architecture and (b) a ViT block. For convolutional layers, kernel size $k$ and output channel $C$ are denoted by ($k \times k$, $C$). For ViT layers, the number of heads $N_h$ and dimensions per head $d_h$ are denoted as ($N_h$, $d_h$). $H \times W \times C$ denotes the height, width, and channels of the image/features, and DS and US stand for spatial downsampling and upsampling, respectively.}
\label{fig:sysarch}
\end{figure*}

\subsection{Multi-Head Self-Attention and Convolutions} \label{conv_vs_mhsa}
Multi-Head Self-Attention (MHSA) mechanism is the essence of the Vision Transformer architecture. It is differentiated from convolution in that it has global receptive fields and content-adaptivity. In this section, we introduce the mathematical formulation of the multi-head self-attention and compare them to convolutions. The notation used is mainly borrowed from~\cite{cordonnier2019relationship}.

\subsubsection{Self-attention}
Let $\mathbf{X} \in \mathbb{R}^{N \times D_{\textrm{in}}}$ represent an input matrix, which has $N$ tokens (or pixels) and $D_{\textrm{in}}$ dimensions. An (single-headed) self-attention block maps an input $\mathbf{X}$ into output $\mathbf{Y} \in\mathbb{R}^{N \times D_{\textrm{out}}}$ dimension output as follows:
\begin{equation}\label{eqn:selfattn}
\mathbf{Y} = \textrm{Self-Attention}(\mathbf{X}) = \textrm{softmax}(\mathbf{A} / \small{\sqrt{D_{\textrm{in}}}} )\mathbf{X}\mathbf{W}_{\textrm{val}}
\end{equation}
where $\mathbf{A}$ is an attention matrix and is defined as:
\begin{equation}\label{eqn:attnmatrix}
\mathbf{A} = \mathbf{X}\mathbf{W}_{\textrm{qry}}\mathbf{W}_{\textrm{key}}^{\mathsf{T}}\mathbf{X}^{\mathsf{T}} + \mathbf{P}
\end{equation}
and $\mathbf{W}_{\textrm{qry}}$, $\mathbf{W}_{\textrm{key}}$, $\mathbf{W}_{\textrm{val}} \in \mathbb{R}^{D_{\textrm{in}} \times D_{\textrm{out}}}$ are query, key, value transformation matricies and $\small{\sqrt{D_{\textrm{in}}}}$ is a normalization factor, respectively. $\mathbf{P} \in \mathbb{R}^{N \times N}$ is a learnable positional encoding that alleviates its permutation-equivariance and translation-variance, which can be problematic for images. Attention matrix $\mathbf{A}$ can be interpreted as a similarity matrix between input tokens in a latent space, and self-attention produces new features by using mutual similarities of given inputs as weights by softmaxing and multiplying the attention matrix.

\par\smallskip
\subsubsection{Multi-head self-attention}
Multi-head self-attention is done by conducting multiple self-attentions (``heads'') in parallel to enable multiple interpretations of the same input exploiting different query, key, and value transformations. It consists of $N_{\textrm{h}}$ heads and $D_{\textrm{h}}$ dimensions per each head, and multiple head outputs are combined to produce final representations as follows:
\begin{equation}\label{eqn:mhsa}
\textrm{MHSA}(\mathbf{X}) = \underset{h \in [N_{\textrm{h}}]}{\textrm{concat}}[\textrm{Self-Attention}_{h}(\mathbf{X})]\mathbf{W}_{\textrm{out}}
\end{equation}
where $\mathbf{W}_{\textrm{out}} \in \mathbb{R}^{D_{\textrm{out}} \times D_{\textrm{out}}}$ is a transformation matrix used for projecting concatenated head outputs, and the output dimension of $\textrm{Self-Attention}_{\textrm{h}}$ is $D_{\textrm{h}}$. Typically $D_{\textrm{\textrm{h}}}$ is set as $D_{\textrm{h}} \times N_{\textrm{h}} = D_{\textrm{out}}$.

\par\smallskip
\subsubsection{Convolutions}
Convolutional layers have been widely adopted in building neural networks for images, even after the advent of self-attentions. For a input feature $\mathbf{X} \in \mathbb{R}^{H \times W \times D_{in}}$, convolutional layers with learned kernel matrix $\mathbf{K} \in \mathbb{R}^{k \times k \times D_{in}}$ can be denoted as follows:
\begin{equation}\label{eqn:conv}
\textrm{Conv}(\mathbf{X})_{i,j} = \sum_{a=\lfloor -{k \over 2} \rfloor}^{\lfloor {k \over 2} \rfloor} \sum_{b=\lfloor -{k \over 2} \rfloor}^{\lfloor {k \over 2} \rfloor}
\mathbf{K}^\mathsf{T}_{\lfloor {k \over 2} \rfloor+a,\lfloor {k \over 2} \rfloor+b,:} \mathbf{X}_{i+a,j+b,:}.
\end{equation}

As seen above, convolutional layers transform given matrices only depending on learned kernels shared by all pixels and are content-agnostic. Their receptive fields, i.e., the range of input pixels utilized to produce a single output token, are limited to the kernel size $k$. In the contrary, multi-head self-attention has a global receptive field and calculates the inner products of the queries and keys, enabling content-dependent operations. However, it requires much more computation and memory resources due to the calculation of $N \times N$ inner products of the attention matrix. It should be carefully used when there are a number of tokens, e.g., high-resolution images.

\subsection{Vision Transformers} \label{intro_vit}

Vision Transformers were first introduced in \cite{ViTpaper}, inspired by a Transformer~\cite{vaswani2017attention}, which is a de facto standard architecture in a natural language processing research community. Unlike CNN, it adopts the multi-head self-attention mechanism to enable content-adaptive operations and has global receptive fields. ViTs and their variants outperform CNNs and record state-of-the-art performances in various computer-vision fields, including image classification~\cite{coatnet}, object detection~\cite{carion2020end}, or image restoration~\cite{zamir2022restormer}.

In the recent research~\cite{ViTprops}, it is found that ViTs are more robust on common image corruptions, such as occlusions, permutations, natural perturbations, or even adversarial attacks. When 80\% of an image is randomly dropped, for example, it shows $\sim$60\% classification accuracy, whereas CNN maintains zero accuracies. It also recorded 36\% lower mean corruption error~\cite{hendrycks2019benchmarking} for natural perturbations, or 30\%p more classification accuracy when an adversarial patch -- 5\% of the total image size -- is added. Also, the authors in \cite{park2022vision} argued that ViTs behave like low-pass filters in image classification, unlike CNNs resembling high-pass filters, which is desirable as low-pass filtering is a typical way to enhance noised images.

\begin{table*}[htbp]
\centering
\caption{Tested architectures, their computational complexity (GFLOPs), trainable parameters, and decoded image quality.}
\begin{threeparttable}
\begin{tabularx}{\textwidth}{>{\hsize=1.2\hsize}X|>{\hsize=.4\hsize}X|>{\hsize=.6\hsize}X|>{\hsize=.6\hsize}X|>{\hsize=.8\hsize}X}
\toprule
\textbf{Architecture} &\textbf{GDN} &\textbf{GFLOPs} &\textbf{Parameters (M)} &\textbf{Image Quality (PSNR)} \\
\midrule
C-C-C-C-C-C (DeepJSCC) &YES	&6.63 &19.9 &30.84	\\
\midrule
\textbf{C-C-V-C-C-C}\tnote{a} &\textbf{YES}	&\textbf{6.44} &\textbf{16.0} &\textbf{31.36}	\\
C-C-V-C-C-C\tnote{a} &NO	&6.33 &15.9 &31.64	\\
C-V-V-C-C-C\tnote{a} &YES	&6.39 &14.9 &31.15	\\
\midrule
\textbf{C-C-C-V-C-C}\tnote{b} &\textbf{YES}	&\textbf{6.54} &\textbf{17.8} &\textbf{31.38}	\\
C-C-C-V-C-C\tnote{b} &NO	&6.43 &17.7 &31.59	\\
C-C-C-V-V-C\tnote{b} &YES	&6.29 &16.5 &30.42	\\
\midrule
C-V-V-V-V-C &YES	&6.05 &11.5 &31.37	\\
C-C-V-V-C-C &YES	&6.35 &13.9 &31.45	\\
\textbf{C-C-V-V-C-C (SemViT, proposed)} & \textbf{NO} & \textbf{6.24} & \textbf{13.8} & \textbf{31.8} \\
\bottomrule
\end{tabularx}
\begin{tablenotes}
\item[a] Only considered the combination of encoder blocks, while keeping the decoders consistent with the baseline configuration.
\item[b] Solely focused on the combination of decoder blocks.
\end{tablenotes}
\end{threeparttable}
\label{tab:combinations}
\end{table*}

Inspired by helpful characteristics of ViTs, such as low-pass filtering effects and robustness to impaired signals, we chose the Vision Transformer architecture and its multi-head self-attention mechanism as our primary building blocks. Our rationale for using ViTs in semantic communications systems is as follows:

\begin{itemize}
\item \textbf{ViTs can perform source coding better than CNNs.}

Vision Transformers have content-adaptivity and global receptive fields, whereas CNNs learn kernel weights and have local receptive fields. Those properties can help source coding, where reducing content redundancies among all features is critical.

\item \textbf{ViTs are more robust to noised signals than CNNs.}

ViTs behave like low-pass filters~\cite{park2022vision} in image classifcation, unlike CNNs, which behave like high-pass filters. As the decoding process can be viewed as denoising, which removes high-frequency artifacts from the channel noises, ViT can outperform CNNs in semantic communications systems. ViTs are also known for their robustness to occlusions or perturbations in images, which resemble channel noises.
\end{itemize}

\section{Semantic ViT} \label{semantic_vit}

\subsection{System Architecture} \label{sysarch}

In this paper, we propose SemViT, an abbreviation for \textbf{Sem}antic \textbf{Vi}sion \textbf{T}ransformer. It follows the typical autoencoder design of a semantic communications system, as described in Section~\ref{intro_semantic}. It has ten blocks, 5 for the encoder and decoder networks, respectively. The encoder network maps input color image $\mathbf{X} \in \mathbb{R}^{H \times W \times 3}$ into complex in- and quadrature-phase messages $\mathbf{M} \in \mathbb{C}^{S}$, and the decoder reconstructs the decoded image $\mathbf{\hat{X}} \in \mathbb{R}^{H \times W \times 3}$ from the channel-corrupted symbols. Following prior works~\cite{bourtsoulatze2019deep, kurka2020deepjscc}, we denote the proportion of the number of complex symbols sent and the number of pixels in the original image, ${S} \over {H\times W\times 3}$, as a bandwidth ratio.

\begin{figure*}[htbp]
\centering{\includegraphics[width=\textwidth]{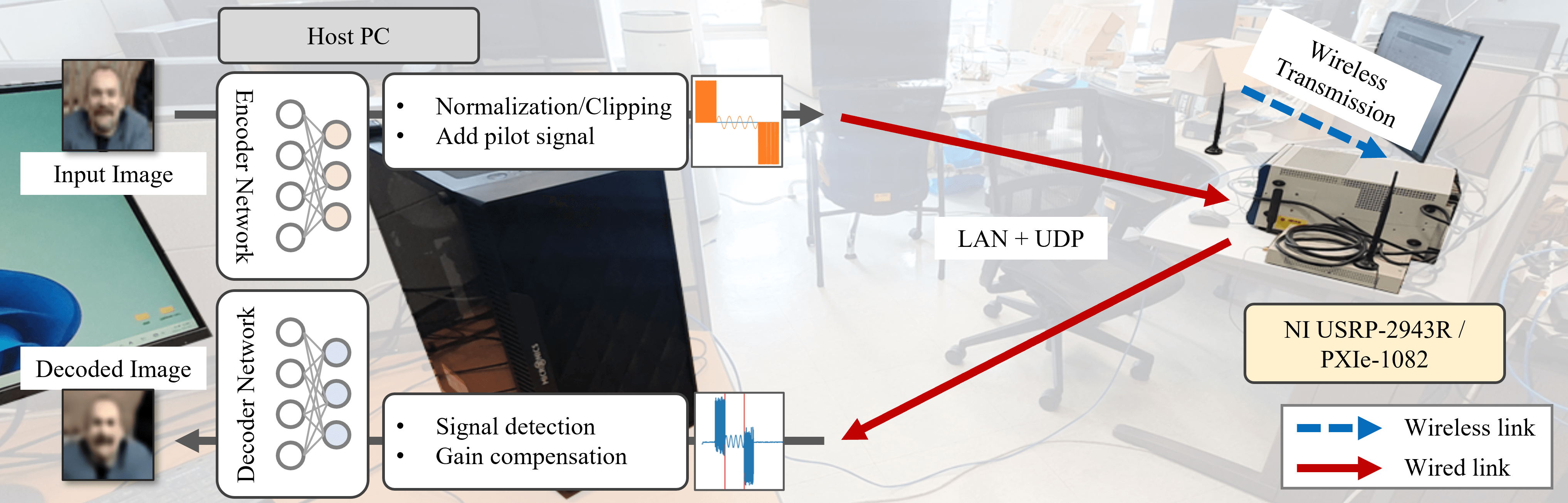}}
\caption{System architecture of the USRP-based wireless semantic communications system testbed.}
\label{fig:usrp_sysarch}

\centering{\includegraphics[width=\textwidth]{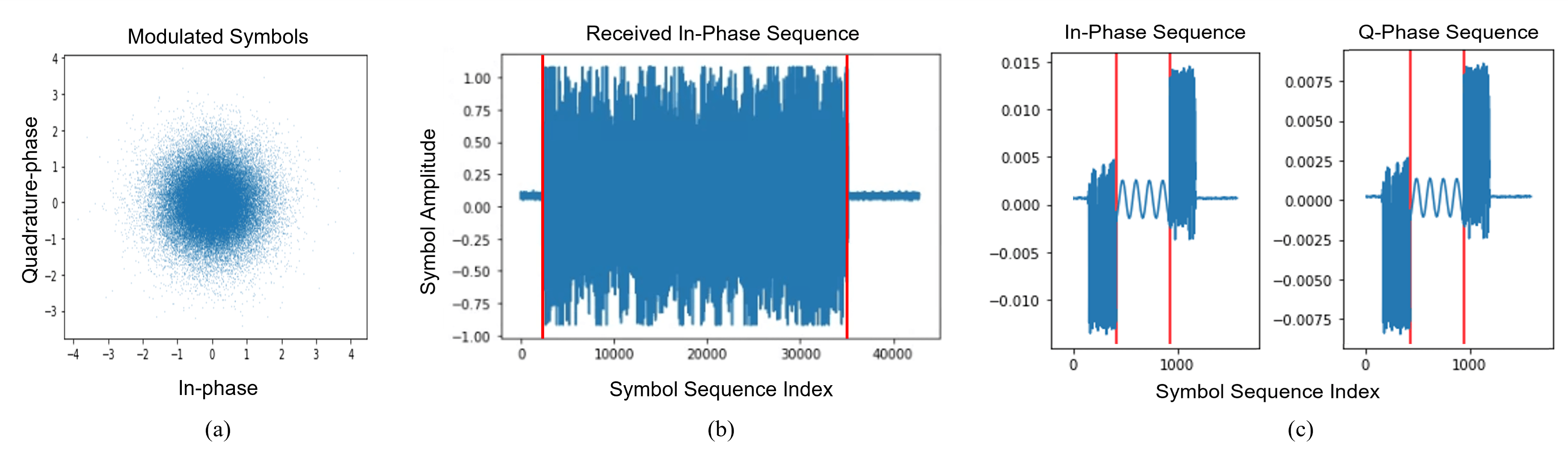}}
\caption{Example I/Q signals transmitted in the wireless testbed. (a) Modulated symbol plot obtained from 64 test images, corresponding to a total of 32,768 symbols. (b) Raw in-phase sequence received in the USRP device. The red vertical lines indicate the estimated symbol duration, obtained by autocorrelation of the pilot sequence added to the front and back of the modulated symbols. (c) Visualization of the I/Q imbalance problem. Despite transmitting only in-phase symbols, the same signal is also received (albeit with weak power) in the quadrature phase.}
\label{fig:proto_signals}
\end{figure*}

As shown in Fig.~\ref{fig:sysarch}, we combine convolutional and ViT layers to build the semantic communications systems. For convolutional layers, we used a kernel size of $9 \times 9$ or $5 \times 5$, following~\cite{kurka2020deepjscc, balle2016end}. We used \cite{ViTpaper} and \cite{coatnet}-inspired ViT layers, which consist of MHSA layers (described in Section~\ref{conv_vs_mhsa}) and multi-layer perceptron layers for ViT blocks. For positional encoding, we used learnable 2D positional encoding $\mathbf{P} \in \mathbb{R}^{N \times N}$ that are sampled from $\mathbb{R}^{(2N-1) \times (2N-1)}$ based on the relative distance between the key and query pixels.

To find the best combinations of the convolution and ViT layers, we started from the CNN baseline (based on~\cite{kurka2020deepjscc, balle2016end}) and replaced one by one each convolutional layer with a ViT layer from the middle of the architecture. We did so because ViTs are known to perform poorly at the first few layers of the neural network (called ``stem")~\cite{coatnet}. To overcome the time-consuming training process associated with testing all possible combinations, we adopted a more efficient approach. Specifically, we replaced convolutional layers selectively either in the encoder (e.g., C-C-V-C-C-C) or the decoder (e.g., C-C-C-V-C-C) to test each combination independently.

We also adhered to the approach suggested in~\cite{coatnet}, imposing a constraint that convolution stages should precede Transformer stages~\cite{coatnet, xiao2021early, wang2018non}. This decision was based on the observation that convolutional layers excel at processing local patterns, which are more prevalent in the early stages of the network. By assessing the performance of these individual combinations, we identified the most effective configurations for both the encoder and decoder parts. Subsequently, we assembled the best combinations to determine the optimal network architecture for the entire system. Considering the time-consuming nature of testing all possible combinations, this approach allowed us to efficiently determine the most effective configurations.

Table~\ref{tab:combinations} shows the tested architecture, computational complexity (GFLOPs), the number of parameters, and the average quality of the reconstructed images (PSNR). We report the image PSNR results with 1/6 bandwidth ratio, 10~dB SNR. GDN denotes whether the generalized divisive normalization layer~\cite{balle2016end} is adopted. We used `C' for convolutional layers and `V' for ViT layers, and their index denotes their layer number. For example, C-C-V-V-C-C means we used ViT layers at layers 2 and 3 (see Fig.~\ref{fig:sysarch}) and convolutional layers for all other layers. Each architecture is trained for 100 epochs.

The C-C-V and V-C-C architecture performed best regarding produced image quality at the encoder and the decoder, respectively. As a result, we adopted the \mbox{C-C-V-V-C-C} architecture to our semantic communications system. Furthermore, this architecture had the lowest computation complexity and trainable parameters, thanks to the parameter-FLOPs efficiency of the ViTs compared to CNNs with large kernel sizes (e.g., 5 or 9). We also found that the generalized divisive normalization (GDN) layer, which is a normalization layer typically used in image compression community~\cite{balle2016end} and prior works~\cite{kurka2020deepjscc}, is not beneficial for ViTs. This is possibly due to the Layer Normalization~\cite{ba2016layer} in the ViT layer, and is also consistent with the recent research result~\cite{bai2022towards}, which reports that GDN resulted in training instability with ViTs. We thus removed it to get additional GFLOPs and parameter reduction. Note that due to hardware architecture and software optimizations, calculated GFLOPs are not proportional to latencies (e.g., the C-C-V-V-C-C architecture is about $1.3\times$~slower at training than is the C-C-C-C-C-C).

\subsection {Training and System Setup} \label{system_setup}
For training and evaluation, we used a desktop PC equipped with Nvidia GeForce RTX 2080 Ti 11~GB. We used the CIFAR-10~\cite{cifar} dataset for training/validations and reported test results based on the CIFAR-100 test dataset, both of which consist of 60,000 $32 \times 32$-pixel color images (50,000 training, 10,000 test images). Unless stated otherwise, we used an Adam optimizer with 0.0001 constant learning rate. We trained the model for 600 epochs.

We implemented a wireless semantic communications system based on the NI USRP-2943R FPGA platform and PXIe-1082 chassis for experiments in a real wireless channel. Fig.~\ref{fig:usrp_sysarch} shows the wireless transmission process of USRP-based semantic communication systems. We first encoded the given image with the encoder network in the host PC. The encoded symbols were then sent to the USRP via LAN with UDP protocol, after which the USRP conducted the wireless transmission of the received symbol. The USRP then delivered the channel-corrupted symbol back to the PC, where the signal was decoded to produce the output image.

\begin{figure*}[htbp]
\centering{\includegraphics[width=\textwidth]{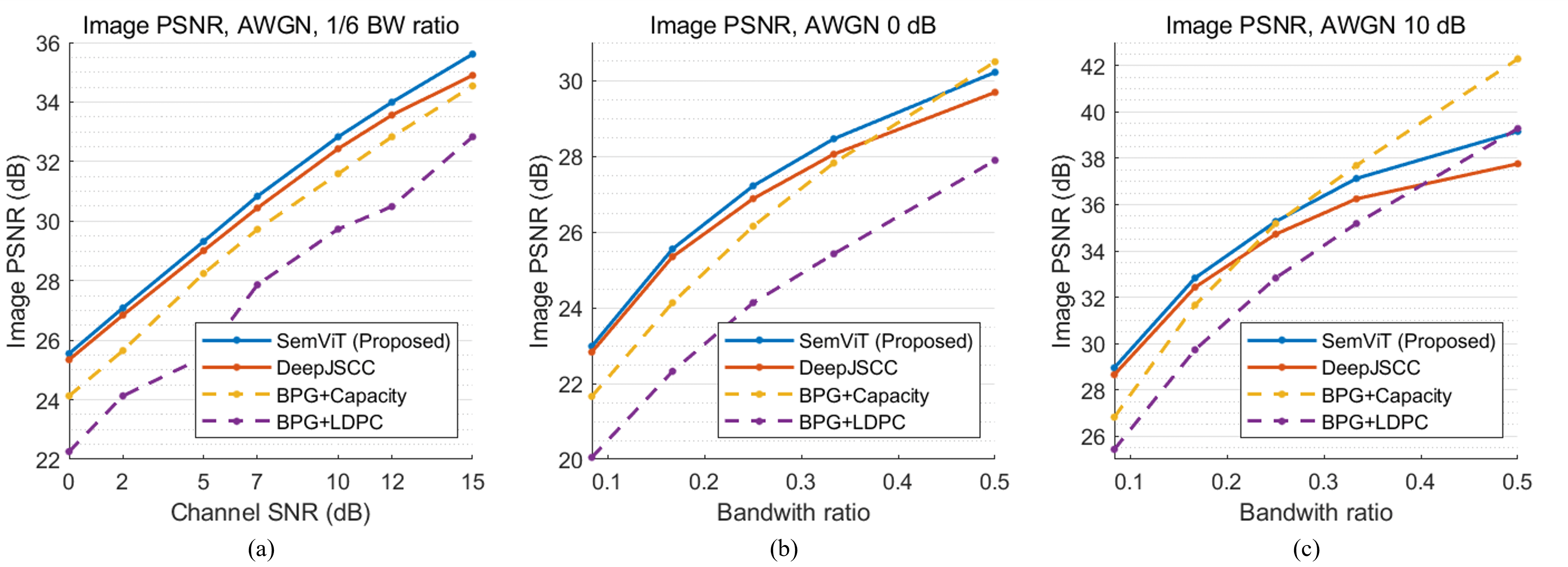}}
\caption{(a): Decoded image quality at bandwidth ratio=$1 / 6$ with respect to channel SNRs. (b), (c): Transmitted image PSNR with regard to bandwidth ratio at 0~dB and 10~dB SNR, respectively. All results are reported from the models trained at the same SNR and BW ratio to the evaluation setup.}
\label{fig:psnr}
\end{figure*}

We conducted wireless transmission experiments using a single USRP device equipped with two omnidirectional antennas: one for transmission and another for reception. The transmission was carried out with a base frequency of 2~GHz and a bandwidth of 1~MHz. The experiments were performed in line-of-sight (LoS) environments, as depicted in Figs.~\ref{fig:usrp_sysarch} and \ref{fig:ces}a. To decode the received signals, we initially performed gain compensation using the pilot signals. Subsequently, we employed a neural network decoder that had been trained in simulated AWGN environments. It is important to note that our focus in this study is to present initial proof-of-concepts for neural network-modulated symbol transmission, and as such, the specific parameters mentioned above can be adjusted as needed.

To address hardware implementation challenges and accommodate the Gaussian-like nature of the neural-network-modulated symbols (as depicted in Fig.~\ref{fig:proto_signals}a), symbol clipping was employed using constant thresholds to align with the digital-to-analog converter (DAC) requirements. Additionally, pilot symbols were introduced at the beginning and end of the transmitted symbols, to enable signal detection via autocorrelation and gain compensation (see Fig.~\ref{fig:proto_signals}b and Fig.~\ref{fig:proto_signals}c).

Fig.~\ref{fig:proto_signals}c reveals the presence of I/Q imbalance problems arising from impairments in our USRP device. To mitigate these issues, we developed a model to characterize the interference between the I/Q symbols, as demonstrated in (\ref{eqn:iq_imbalance}). In these equations, $\mathbf{\hat{x}}_{i}$/$\mathbf{\hat{x}}_{q}$ represents the interfered I/Q symbols, $\mathbf{x}_{i}$/$\mathbf{x}_{q}$ denotes the original I/Q sequence, and $k_{i}$/$k_{q}$ represents the interference constant associated with each respective component:
\begin{align}
\label{eqn:iq_imbalance}
\mathbf{\hat{x}}_{i} =  \mathbf{x}_{i} + k_{q} \mathbf{x}_{q}, \\
\mathbf{\hat{x}}_{q} =  k_{i} \mathbf{x}_{i} + \mathbf{x}_{q}.
\end{align}
In the next step, we estimated the constants $k_{i}$ and $k_{q}$ by utilizing the pilot signal transmitted prior to the main symbol transmission. Through this calibration process, we successfully obtained the I/Q symbols without interference. It is important to note that these calibration procedures, as well as our coarse hardware implementations, introduced additional noise. More precise hardware implementations have the potential to further enhance performance, and we leave them for future work.

\begin{figure*}[htbp]
\centering{\includegraphics[width=\textwidth]{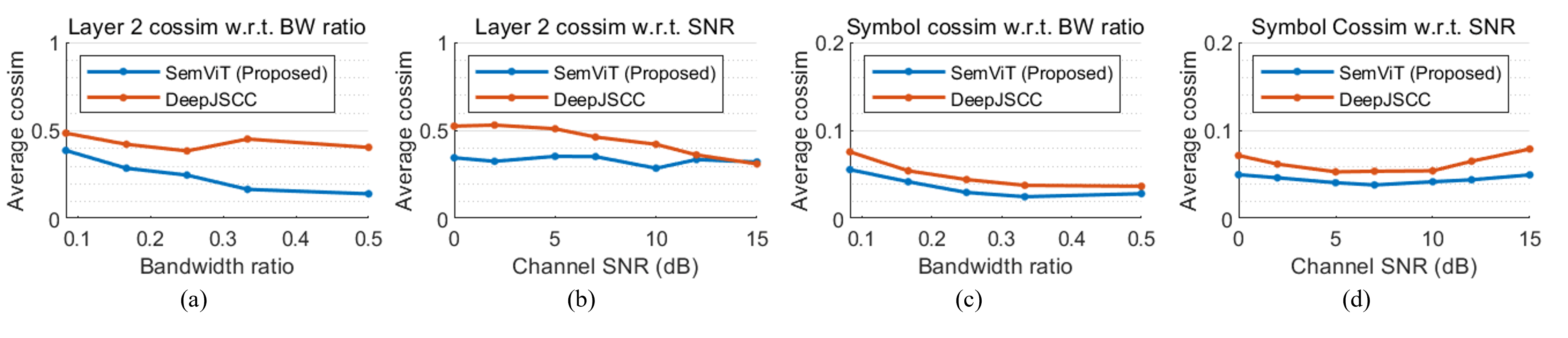}}
\caption{(a), (b): Cosine similarity at layer 2 (last features before symbol projection layer) with respect to bandwidth ratio and channel SNR, respectively. (c), (d): Cosine similarity at complex symbol with respect to BW ratio and SNR. (a) and (c) are measured in AWGN 10~dB channel, and (b) and (d) are with 512~symbols. Note that layer two output has 256-dims while the complex symbol is a 2-dimensional vector, so a direct comparison of cossim values might be unfair. }
\label{fig:cossim}

\centering{\includegraphics[width=\textwidth]{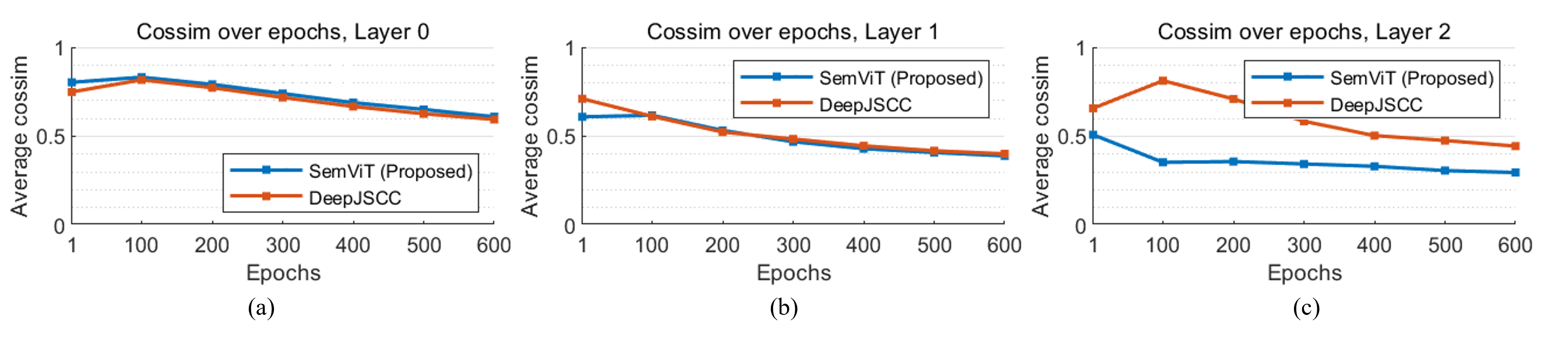}}
\caption{Evolution of average cosine similarities over epochs in encoder network. Every encoder layer produces features with lower average cosine similarities as the training continues. (a), (b), (c): Cosine similarities at layers 0, 1, and 2, respectively. See Fig.~\ref{fig:sysarch} for layer naming.}
\label{fig:cossim_epochs}
\end{figure*}

\section{Results and Analysis} \label{results}

\subsection {Image PSNR performance} \label{psnr_performance}

Fig.~\ref{fig:psnr} compares decoded image quality (PSNR) between proposed SemViT, CNN-based DeepJSCC, and conventional better portable graphics\footnote{BPG format is an image format based on HEVC (H.265) video codec and is one of the most efficient image compression methods among the non-neural network-based image codec.} (BPG)~\cite{bpgformat} format. To facilitate a direct comparison with previous works~\cite{kurka2020deepjscc}, we considered several bandwidth ratios, namely 1/12, 1/6, 1/4, 1/3, and 1/2. These ratios were chosen to align with the parameters used in the referenced study, allowing for easy and meaningful comparisons between our results and those reported in the literature.

For the BPG+LDPC approach, we evaluated the performance in terms of the best PSNR value across various combinations of code rates and modulations. We evaluated various combinations of code rates and modulations, including (3072, 6144), (3072, 4608), and (1536, 4608), which correspond to 1/2, 2/3, and 1/3 code rates, respectively. In terms of modulations, we considered options such as BPSK, 4-QAM, 16-QAM, or 64-QAM. To ensure compatibility between the code bits and the bits per symbol in the selected modulation schemes, we made sure that the number of code bits was divisible by the number of bits per symbol (1, 2, 4, or 6). On the other hand, for the BPG+capacity approach, we focused on calculating the theoretical capacity of the AWGN channel based on the SNR, which was expressed in bit/s/Hz or bit/symbol. Using this information, we determined the required image file size that matched the number of symbols transmitted. Consequently, we adjusted the quality factor of BPG compression to ensure compliance with the file size restrictions. Finally, we reported the PSNR of the encoded image obtained using this approach.

As expected, the proposed SemViT outperformed DeepJSCCs in all regions, proving ViT's benefits in semantic communications. Interestingly, the performance gap between SemViT and DeepJSCC increased as the channel SNR and bandwidth ratio rose, effectively narrowing the gap between conventional separate source-channel coding-based methods and semantic communications in the high SNR and bandwidth ratio region. DeepJSCC particularly underperformed BPG+capacity at AWGN 10 dB, 1/4 bandwidth ratio (Fig.~\ref{fig:psnr}c), but SemViT effectively utilized given data rates and SNR to beat the DeepJSCC and even BPG+capacity.

This is in accord with our first intuitions--\textbf{SemViT can effectively reduce redundancies between representations and diversify output features}, thanks to its content-adaptiveness and global receptive field. According to this interpretation, the ViT's ability to produce diverse features led to a more significant performance gap in the higher SNR and bandwidth ratio region, where semantic communication systems should convey more detailed information (e.g., high-frequency components of the image). The lower gap in the lower SNR and bandwidth region can be explained by the CNN's ability to extract critical features in the image or the necessity of redundant features to deal with the harsher channels. To support these ideas, we chose two metrics for the analysis--cosine similarity and Fourier analysis. Detailed analysis and rationales are explained in the following sections.

\subsection {Cosine similarity analysis} \label{cossim_analysis}
We chose a metric spatial-wise average cosine similarity to show the diversity of the produced features. We interpreted each layer's output features $\mathbf{X} \in \mathbb{R}^{H \times W \times C}$ as a set of $C$-dimensional vectors and computed average cosine similarities $S$ between all $H \times W$ vectors as follows:

\begin{equation}\label{eqn:attnmatrix}
S = \frac{1}{HW(HW-1)} \sum_{i,j} \sum_{\substack{p,q \\ (p,q) \neq (i,j)}} \frac{\mathbf{X}_{i,j,:}^{\mathsf{T}} \mathbf{X}_{p,q,:}}{\left\|\mathbf{X}_{i,j,:}\right\|\left\|\mathbf{X}_{p,q,:}\right\|}.
\end{equation}

\par\smallskip
\noindent\textbf{SemViT produces more diverse features.} In Fig.~\ref{fig:cossim}a, we show the average cosine similarity with respect to the number of symbols in layer 2, which produces final features before being projected to symbols. SemViT consistently reduced the average cossims as the number of symbols increased, while DeepJSCC failed to reduce them in a large number of symbol regions. Instead, in DeepJSCC, symbol diversification is conducted by the mere weighted sum of the features in the linear projection layer behind (see Fig.~\ref{fig:cossim}a and Fig.~\ref{fig:cossim}c). This coincides with the tendency for the image-quality gap to increase as the number of symbols increases. Furthermore, among all regions, the average cosine similarity of the SemViT is significantly lower than DeepJSCC. Considering both results, we can interpret the performance gain as a result of SemViT's ability to produce more diverse features, thanks to its content-adaptiveness and global receptive field. CNN's failure to produce diverse features can be explained by the significant redundancy of learned filters~\cite{denil2013predicting} (a well-known problem), leading to redundant representations.

Note that cosine similarity was not incorporated as a metric in our loss function during the training of the neural network, and the cosine similarity results are obtained from the network's primary objective of maximizing its decoding performance. Our insights based on cosine similarity analysis are as follows: 

\begin{figure*}[htbp]
\centering{\includegraphics[width=\textwidth]{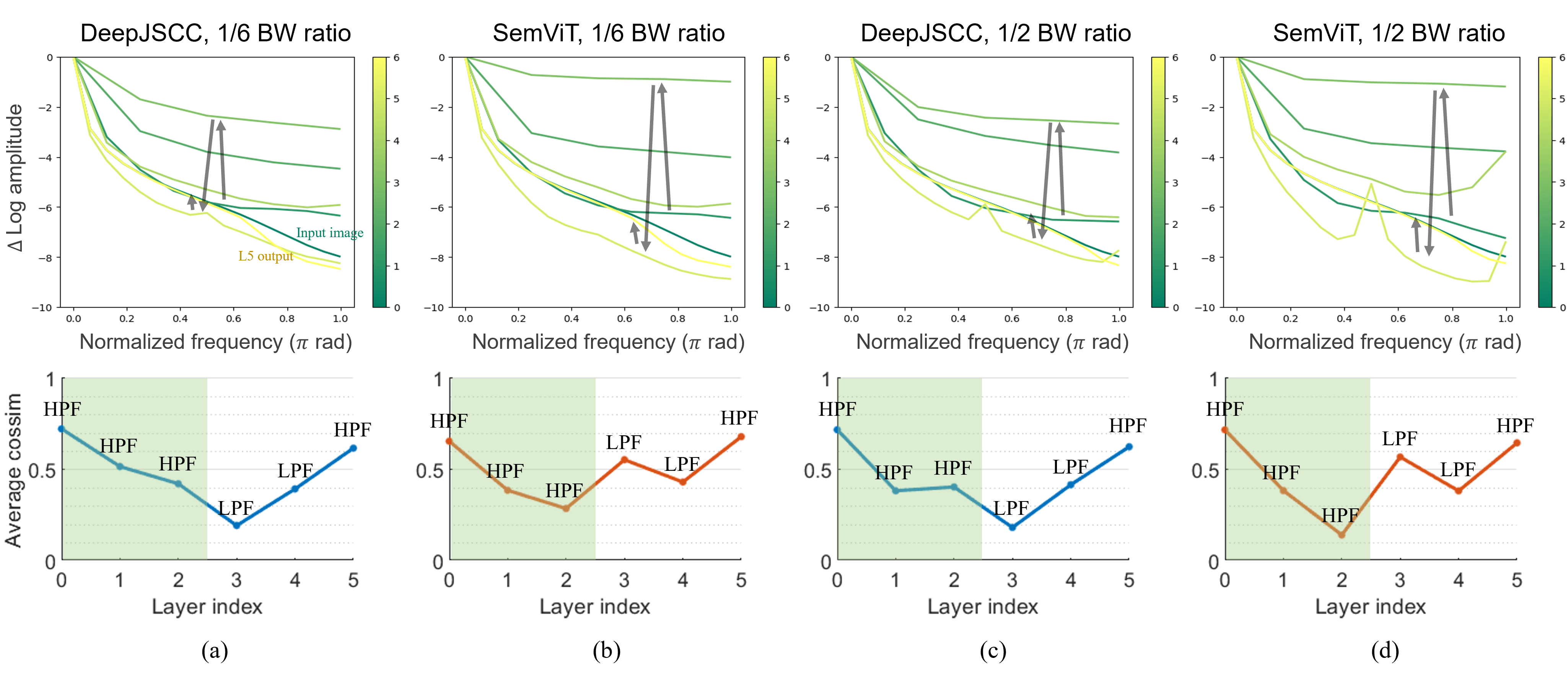}}
\caption{top: fourier analysis from the input image to the final layer output. The gray arrow shows how the relative amplitude changes as the layer index increases. Bottom: Layer-wise average cosine similarity. We denoted whether the given layer is a high-pass filter (HPF) or a low-pass filter (LPF) based on the fourier analysis. The colored region (layers 0-2) represents the encoder part.}
\label{fig:fourier}
\end{figure*}

\par\smallskip
\noindent\textbf{ViTs can be good at source coding, while CNNs may be so at channel coding.}
In Fig.~\ref{fig:cossim}b, DeepJSCC shows a weak tendency to diversify its output features as the channel SNR rises, whereas SemViT shows no significant tendency to do so. SemViT's invariability to channel SNRs is rather complemented by the linear projection layer behind (see Fig.~\ref{fig:cossim}b and Fig.~\ref{fig:cossim}d). This implies the specialization between ViTs (for the source coding) and projection layers (kind of channel coding). In contrast, CNN's cosine similarity at layer 2 is inversely proportional to channel SNRs, while the tendency becomes weaker in the symbols produced by the 1D-projection layer (See Fig.~\ref{fig:cossim}b and Fig.~\ref{fig:cossim}d). This pattern suggests the possibility that the CNNs are best at the channel coding among the ViTs, projection layers, and CNNs. Also, the inconsistency of symbol cosine similarity with respect to train SNRs may be due to oversimplified symbol production. As a solution for building symbols, linear projection followed by reshaping might not be a good one. Note that the architecture of our proposed SemViT is not based on these insights, as our purpose in designing SemViT is to find initial designs of ViT-based semantic communications systems and to provide preliminary analysis. We leave the architecture design based on those understanding to future work.

\par\smallskip
\noindent\textbf{Appropriateness of the metric.}
Judging the effectiveness of the chosen metric is a difficult work. Hence, we show that the network is trained to reduce average cosine similarity. Fig.~\ref{fig:cossim_epochs} shows the average cosine similarity of the encoder layers decreases as the training epochs lengthen, showing the association between lower cosine similarities and the image quality. Furthermore, as shown in Fig.~\ref{fig:cossim}a, the cosine similarity gap between SemViT and DeepJSCC increases as the channel SNR and bandwidth ratio rises, which aligns with the PSNR results. We are not arguing that average cosine similarity is the perfect metric that measures the amount of information, but we believe it is good enough to show the difference between the ViTs and CNNs.

\subsection {Fourier analysis} \label{fourier_analysis}
We perform a Fourier analysis on the input image, analyzing the behavior of each layer's frequency filtering up to the final feature prior to image generation (i.e., the output of layer 3). Specifically, we begin by averaging the features produced by layer 2 across the channel dimension to obtain a matrix $\mathbf{X} \in \mathbb{R}^{H \times W}$, and then calculate the log-amplitude difference from the DC component.To maintain clarity and consistency with prior work~\cite{park2022vision}, we report only the half-diagonal portion of the 2D discrete Fourier transform (DFT) of the averaged features. The mathematical formula for obtaining the Fourier analysis result $\mathbf{y} \in \mathbb{R}^{\lfloor \frac{H}{2} \rfloor + 1}$ of the given features, computed from 2D DFT representation $\mathbf{F}$ of the given features $\mathbf{X}$,  is as follows:
\begin{equation}\label{eqn:2d_dft}
\mathbf{F}_{k,l} = \sum_{n=0}^{H-1} \sum_{m=0}^{W-1}{\mathbf{X}}_{n,m}\mathnormal{e}^{-j2\pi \frac{nk}{H} + \frac{ml}{W}},
\end{equation}
\begin{equation}\label{eqn:log_amp_and_diag}
\mathbf{\hat{y}} = \log (\textrm{diag}(|\mathbf{F}|)) - \log (\textrm{diag}(|\mathbf{F}_{0,0}| \mathbf{I})),
\end{equation}
\begin{equation}\label{eqn:half-diag}
\mathbf{y} = \mathbf{\hat{y}}_{0:\lfloor \frac{H}{2} \rfloor + 1}.
\end{equation}
Note that the operator $\textrm{diag}(\cdot)$ generates the diagonal vector from a given matrix, while $\textrm{log}(\cdot)$ carries out an element-wise logarithmic operation. Additionally, $|\mathbf{F}|$ indicates element-wise absolute value of complex-valued matrix $\mathbf{F}$, and $|\mathbf{F}_{0,0}|$ signifies the magnitude of the DC component. Equation \ref{eqn:half-diag} calculates the half-diagonal elements from the computed difference in log-amplitude. Based on the Fourier analysis, we have observed the following results:

\begin{figure*}[htbp]
\centering{\includegraphics[width=\textwidth]{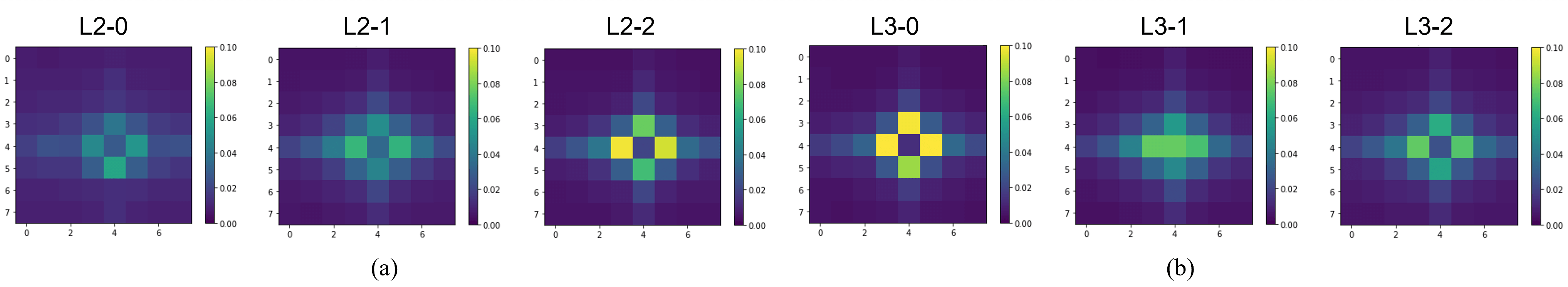}}
\caption{Visualization of sublayer attention map at layers 2 and 3 on the index (4, 4). The symmetrical structure of global-to-local attention is clearly visible.}
\label{fig:attmap}
\end{figure*}

\par\smallskip
\noindent\textbf{Encoders are high-pass filters, while decoders are low-pass filters.}
As can be seen in Fig.~\ref{fig:fourier}, in the encoder network, the amplitude of high-frequency components is continuously increased, whereas the decoder works oppositely. Layer~5 exceptionally acts as a weak high-pass filter, possibly due to the generation of high-frequency details of the image, but the relative amplitude difference is not that significant. Interestingly, unlike CNNs that behave like HPFs in the image classification tasks~\cite{park2022vision}, convolutional layers in the decoder network consistently behave like LPFs. This is likely due to the ``unpacking'' properties of the decoder, which decodes the highly compressed (high-pass filtered) features produced by encoder networks. Also note that the operating dynamics of the CNN can be different between the image classification network and the semantic communications system, as the training objective is different.

\par\smallskip
\noindent\textbf{ViTs behave like strong LPFs in the decoder.}
This affirms that ViT's usage in the decoder network is a good idea, as the decoders are essentially low-pass filters (possibly to suppress the high-frequency noise induced by the channel). Fig.~\ref{fig:fourier} shows the relative amplitude differences of layer 3, in the high-frequency regime, are significant in SemViT compared to DeepJSCC. Interestingly, although layer 3 acts like low-pass filters, the cosine similarity decreases in DeepJSCC; This can be due to the weak low-pass filtering effects of DeepJSCC and the extraction of high-dimensional features (256-dims) from the extremely low-dimensional symbol vectors (e.g., for 1/6 BW ratio, reshaped dimension of layer 3's input is 8). Also, layer 4 in SemViT decreases cosine similarity while acting like a low-pass filter, which can be because of selective solid amplification of mid-band signal (around $0.5\pi$ radians, see Fig.~\ref{fig:fourier}d) while inhibiting other high-frequency regions.

\begin{figure*}[htbp]
\centering{\includegraphics[width=\textwidth]{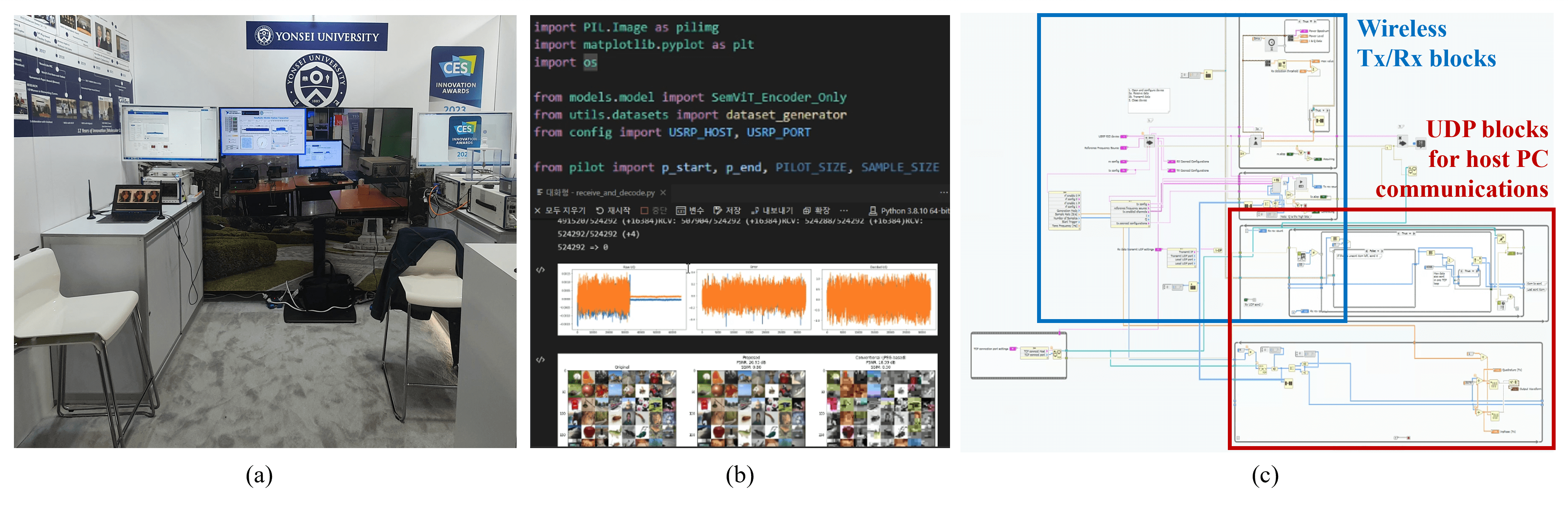}}
\caption{(a):~Real-time demonstration of the proposed system in a crowded indoor environment (at CES 2023, Las Vegas, USA), (b) screenshots of the client/server (for neural encoding/decoding), and (c) the LabVIEW block diagram (for real-time wireless transmission). Both softwares, including neural network parameters, are available open-source. See https://bit.ly/SemViT/.}
\label{fig:ces}

\centering{\includegraphics[width=\textwidth]{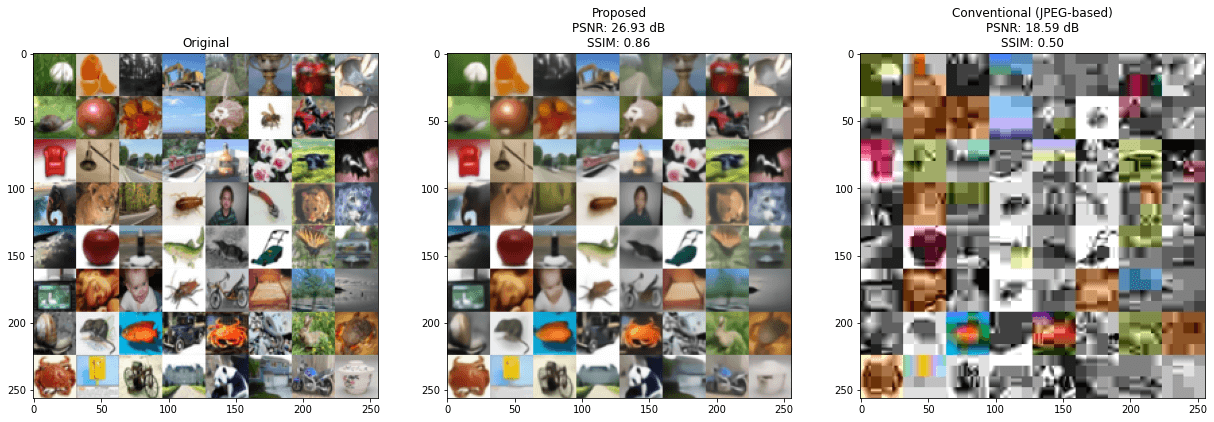}}
\caption{Examples of original (left) and transmitted images using proposed SemViT (middle) and conventional JPEG (right). For the JPEG image, we assumed LTE's modulation-and-constellation scheme targetting 0~dB SNR and inversely calculated the required image bytes to send an equal number of symbols compared to our proposed system. Transmitted images are randomly chosen from the CIFAR-10 test set.}
\label{fig:image_examples}
\end{figure*}

\par\smallskip
\noindent\textbf{ViTs produce many more high-frequency details, especially in the high-BW region.}
Comparing Fig.~\ref{fig:fourier}a and Fig.~\ref{fig:fourier}c, or Fig.~\ref{fig:fourier}b and Fig.~\ref{fig:fourier}d suggests the key difference between the low- and high-BW ratio region is the amplitude of the high-frequency components. Generally, the high-frequency components of the image contain the details (e.g., textures), while the low-frequency parts consist of the rough shape. Therefore, we can interpret the PSNR gain on the high-BW ratio in Fig.~\ref{fig:psnr} is thanks to the increased details. Furthermore, SemViT shows a much larger amplitude in mid- and high-frequency regions than DeepJSCC, which are also compliant with the more significant PSNR gap at a larger BW ratio (Fig.~\ref{fig:psnr}c) and the conclusion of Section~\ref{cossim_analysis}. Note that the ViT decoder still behaves like a robust low-pass filter (i.e., the overall amplitude difference of the final encoder output and the ViT decoder features are relatively huge in high-frequency regions), but selectively retains high amplitude in certain high-frequency bands (e.g., $0.5\pi$ and $1\pi$)

\subsection {Attention maps} \label{att_maps}
In Fig.~\ref{fig:attmap}, we visualize the sublayer attention map of layer~2 (the last layer before the symbol projection) and layer 3 (the first layer of the decoder). For visibility, we chose the attention map at the index (4, 4) and reshaped it to match the feature map size ($8 \times 8$). The results are averaged over all CIFAR-10 test sets (10,000 images).

Surprisingly, the attention map clearly shows the symmetric structure of global-to-local attention; This resembles the context, global- and local- hyperprior structure, which was recently proposed by deep-learning-based image compression community~\cite{kim2022joint}. This might be further evidence showing ViT's strength in source coding. The evident cross-structure of the attention map is due to the additive positional encoding (not shown in the paper). The low self-weighting is likely due to the residual connection of the model architecture (i.e., the previous features are added to the next feature even without the attention procedure).

\section {Results in Other Environments} \label{other_channel_and_metric}

To see the generalization of the results in other channel environments or metrics, we additionally show the results in the Rayleigh fading channel, in the real wireless channel, and with the structural similarity index metric (SSIM)~\cite{wang2004image}, which is a perceptual image quality metric.

\begin{figure*}[htbp]
\centering{\includegraphics[width=\textwidth]{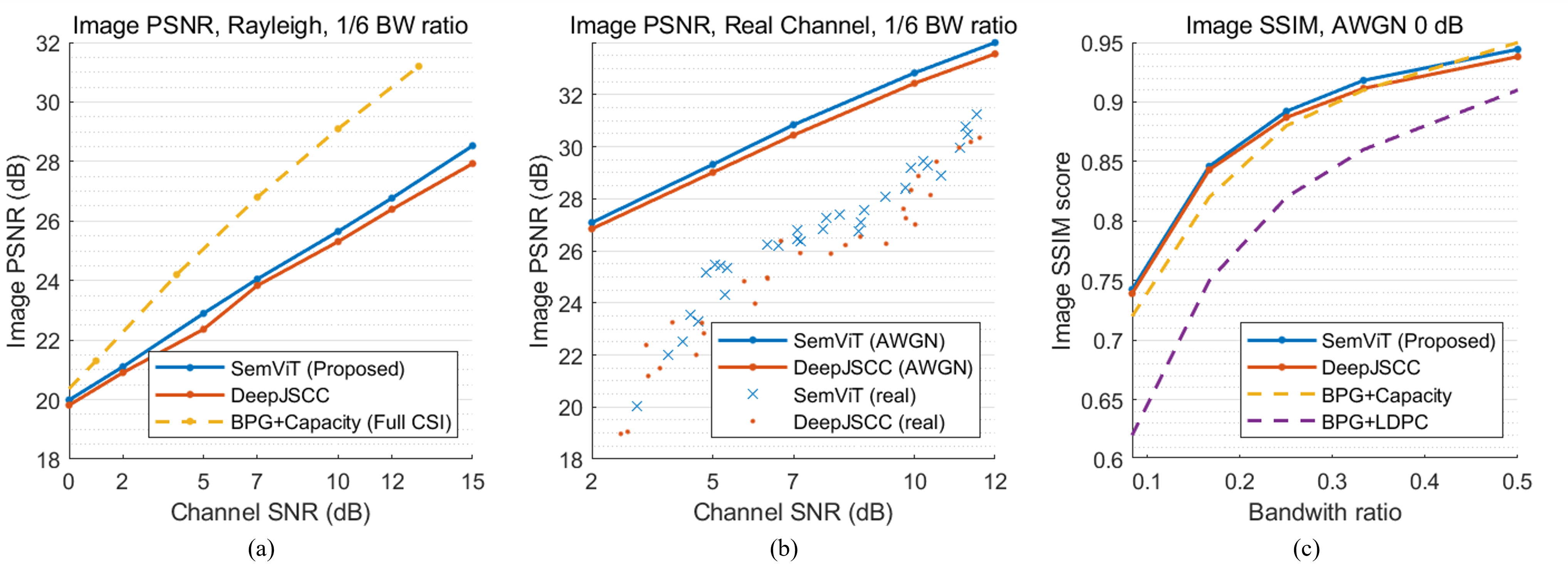}}
\caption{(a), (b): PSNR results in the Rayleigh and real wireless channel, respectively. (c): Image SSIM results in the AWGN, 0~dB SNR. We borrow the BPG+capacity data in (a) from~\cite{kurka2020deepjscc}.}
\label{fig:other_envs}

\centering{\includegraphics[width=\textwidth]{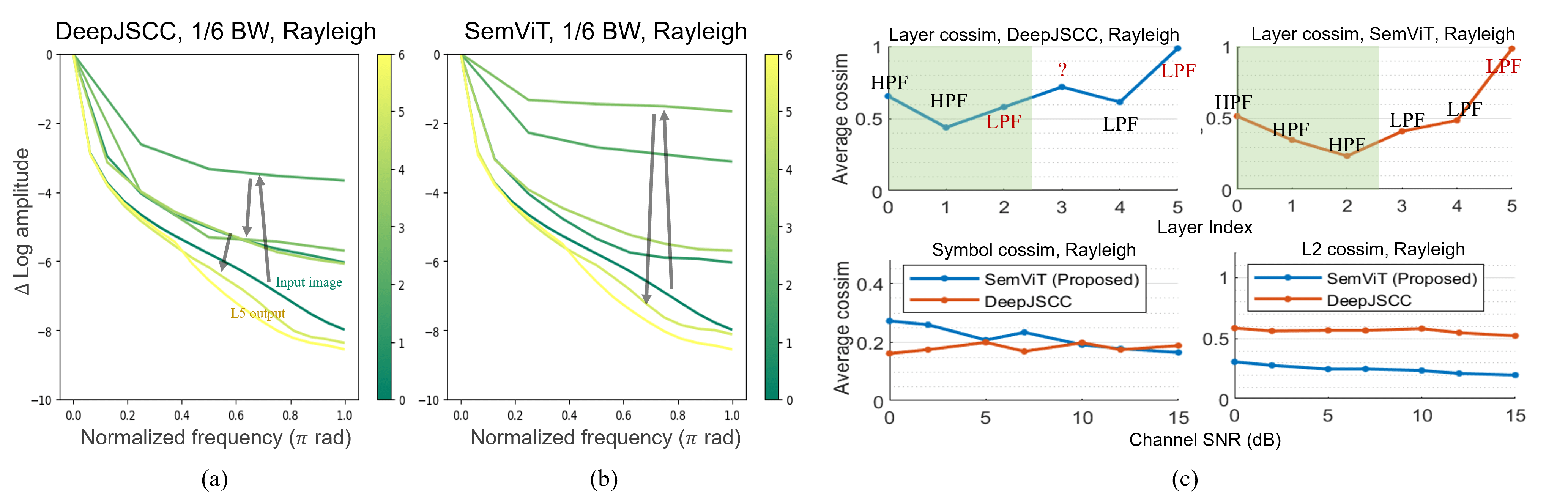}}
\caption{(a), (b): Fourier and (c): cosine similarity analysis in Rayleigh fading channels.}
\label{fig:rayleigh_results}
\end{figure*}

\subsection{Decoded image quality} \label{other_env_quality}

Fig.~\ref{fig:other_envs}a shows the results in the slow Rayleigh fading channel, where the channel is kept unchanged during the transmission of the entire image. We retrained the entire network to match the Rayleigh fading environment without any channel state information to the network. SemViT still outperforms DeepJSCC, especially in the high SNR regions. The PSNR gap between DeepJSCC and the proposed SemViT increases as the SNR rises, coinciding with the AWGN results and our previous analysis.

Fig.~\ref{fig:other_envs}b shows the PSNR results in the real wireless channel, which are measured in the USRP-based prototype mentioned in Section~\ref{system_setup}. We transmitted encoded constellations at different SNR regions by adjusting the amplitude of the transmitted signals, i.e., dividing the encoded signal with varying constants. Each points in the figure denote the average image PSNR of the transmission of the 64 random images. As expected, the proposed SemViT shows better image quality than DeepJSCC in the real wireless channel. However, there exists about a 3 dB PSNR gap between the simulated (AWGN) results and actual measurements due to non-Gaussian errors of the channel (i.e., errors induced by imperfect gain compensations, reflections, and quantization errors of the DAC). We further validated the system in the crowded indoor environment at CES 2023 (Fig.~\ref{fig:ces}), and confirmed there was no significant difference in the performance (Fig.~\ref{fig:image_examples}).

Interestingly, the performance gap between the simulated and real environment rises as the channel SNR degrades. This is likely due to the disparity in the precision between the simulation (32-bit floating points) and the USRP hardware (12-bit fixed-point DAC hardware). As we manipulated the signal power (and the resulting channel SNR) by reducing the signal amplitude, more quantization errors were induced in the lower SNRs (due to the fixed-point arithmetic of DAC hadware) to degenerate the reported image quality. This suggests that the semantic communications system should be trained to consider RF hardware restrictions (e.g., DAC quantizations, power-to-average-power-ratio (PAPR) constraints). We leave them for future work.

Fig.~\ref{fig:other_envs}c reports the SSIM score of the proposed SemViT and DeepJSCC in the AWGN channel. Unsurprisingly, the proposed SemViT shows better SSIM scores compared to DeepJSCC, which are coincident with image PSNR results (Fig.~\ref{fig:psnr}b). Note that the reported SSIM values are with the model trained with PSNR loss and thus can be improved by using the SSIM loss in the training procedure. More perceptual loss, e.g., MS-SSIM~\cite{wang2003multiscale} or VGG loss~\cite{johnson2016perceptual}, might be utilized to enable more semantical compression of the images.

\subsection{Analysis in the Rayleigh fading channel} \label{rayleigh_analysis}
To see if our analysis in the AWGN channel can be generalized to the Rayleigh fading channel, we conducted the Fourier and average spatial-wise cosine similarity analysis in the Rayleigh channel. Fig.~\ref{fig:rayleigh_results}a and Fig.~\ref{fig:rayleigh_results}b shows the Fourier analysis results. The filter characteristics of the encoder and the decoder in AWGN and Rayleigh channel are almost identical. The proposed SemViT trained in the Rayleigh channel behaves like a more robust low- or high-pass filter, coinciding with the previous analysis. However, the amplitude deviation of encoded features is relatively minor in the Rayleigh channel in both DeepJSCC and SemViT, i.e., the encoder network produces much less diverse features to deal with the more unpredictable channel. Also, layer 3 in DeepJSCC does not conduct notable transformations in frequency domains, whereas SemViT still shows obvious LPF behavior.

Notably, DeepJSCC produces low-pass-filtered features in layer two, in which SemViT in the Rayleigh channel or even DeepJSCC in the AWGN channel conducts high-pass filtering (top left in Fig.~\ref{fig:rayleigh_results}c). Due to the CNN decoder's weakness in low-pass filtering, or decoding, DeepJSCC should produce more redundant features in the encoder (LPF behavior) rather than dimensionality reduction (HPF behavior) as like in AWGN channel or SemViT, leading to poorer decoded image quality (Fig.~\ref{fig:other_envs}a). This phenomenon is also seen in cosine similarity analysis results -- in Rayleigh channels, DeepJSCC's average cosine similarity of layer 2 output increases (top left of Fig.~\ref{fig:other_envs}c) whereas SemViT (top right) or DeepJSCC in AWGN (Fig.~\ref{fig:fourier}a) lowers the feature cosine similarity in layer 2. DeepJSCC and SemViT conduct low-pass filtering in the last layer (layer 5), leading to more amplitude differences in high-frequency regions between the original image and the layer five output. This explains the lower image PSNR quality in the Rayleigh channel.

The top part of Fig.~\ref{fig:rayleigh_results}c shows the layer-wise cosine similarity and whether the layer is HPF or LPF, based on the Fourier analysis. The key difference compared to AWGN results is that the cosine similarity of the last layer's features almost equals 1 in both DeepJSCC and SemViT. One possible interpretation is that the semantic communication systems try to average all given symbols spatially to deal with harsh channel conditions and produce a single polished vector from which the network reconstructs the entire image. The DeepJSCC decoder's consistent increase of the cosine similarity, both in Rayleigh and AWGN (Fig.~\ref{fig:rayleigh_results}c and Fig.~\ref{fig:fourier}a), can be explained well in this way. However, layer 4 of AWGN-trained SemViT behaves differently from the DeepJSCC or Rayleigh-trained SemViT -- it diversifies its output features while performing low-pass filtering if the channel is good enough (e.g., AWGN 10~dB SNR as in (Fig.~\ref{fig:fourier}b). Considering the content-adaptivity of the ViTs, this may mean that SemViT decodes the signal by aggregating similar information across all symbols to produce multiple ``pure'' information sources, while DeepJSCC can only average the adjacent symbols due to its content-agnostic property. This suggests that using ViTs may also benefit interference-canceling applications, e.g., inter-symbol interference or self-interference cancellation in full-duplex communications.

In Fig.~\ref{fig:rayleigh_results}, the bottom left illustrates the 2D symbol, while the bottom right shows the layer two feature cosine similarity analysis. As in Fig.~\ref{fig:cossim}b, the layer two cosine similarity of DeepJSCC gradually decreases as the channel SNR rises, i.e., encoded features has some channel adaptivity. Even SemViT, whose features were channel-agnostic in AWGN, produces more diverse symbols as the channel improves. This could be because the Rayleigh channel was so harsh that even the channel-insensitive ViT had to adapt to the SNR, or decoding the symbols in AWGN might be too easy for ViTs. The cosine similarity of the 2D symbols does not show any evident tendencies to adapt better to channel environments, which is corresponsive to AWGN results (Fig.~\ref{fig:cossim}d), and still suggests using a more robust network for final symbol production.

\section {Discussions} \label{discussions}
In this section, we provide some insights and possible research directions based on the analysis given in Section~\ref{results}.

\begin{itemize}

\item Using heterogeneous architectures for semantic communications, which combine the strengths of both ViTs and CNNs, may be more beneficial than relying on either approach alone. This could involve using ViTs for source coding and CNNs for channel coding, with possible applications including the extraction of hierarchical features and the creation of more robust models that are resistant to channel noise.

\item ViTs can serve as effective LPFs that aid in decoding. To test their impact on performance, we can incorporate at least one ViT layer into the decoder network or introduce a non-trainable, explicit blurring layer prior to the decoder network and analyze the results.
\item A more robust network may be necessary for symbol-producing layers. Although we did not observe any clear inverse-proportional relationships between channel SNR and symbol cosine similarities, this may be attributed to the current oversimplified symbol projection layer. To address this, future research may explore the use of channel-wise attention to generate symbols instead of the simple projection and reshaping method.


\item More efficient ViT-based models can be developed to reduce encoding/decoding latencies. Our work did not specifically focus on the computational efficiency of SemViT or compare it to conventional image transmission systems. This is because conventional communication systems typically operate on specialized hardware, whereas our proposed system utilizes a general-purpose graphics processing unit, making a fair comparison difficult. However, with the critical latency requirements of future B5G/6G systems in mind, it is imperative to research more efficient neural network architectures for semantic communications.
\end{itemize}

\section{Conclusion} \label{conclusion}
The SemViT system proposed in this paper used a Vision Transformer to enhance image transmission performance in semantic communications. The experiments conducted in various regions show that SemViT outperforms conventional CNN-based methods in all regions, particularly in high-SNR and bandwidth ratio regimes. We verified the system's availability in real-world wireless channels by conducting extensive experiments on a USRP-based wireless semantic communications testbed, which have been made publicily available as an open-source to enable reproducibility. We also conducted a thorough analysis to determine how a Vision Transformer can improve semantic communications systems. Our analysis stated that
1) encoders are essentially HPF and decoders are LPF, 2) ViTs are good source-coders and diversify the encoded representations, 3) ViTs are beneficial in decoders thanks to their strong LPF behavior, and 4) CNN might be good at channel coding, affirming the combined usage of ViT and convolutional layers. We hope our work provides some deeper insights and facilitates further studies.


\section*{Acknowledgement}
We are grateful to T. Jung for his valuable advice and help in implementing our semantic communications testbed. 

This work was partly supported by the Institute of Information \& communications Technology Planning \& Evaluation (IITP) and the National Research Foundation of Korea (NRF) grant funded by the Korean government (MSIT) (No. 2021-0-02208, 2022R1A5A1027646) and Samsung Electronics.

\bibliographystyle{IEEEtran}
\bibliography{ref}

\end{document}